\renewcommand{\Im}{\mathop{\rm Im}\nolimits}
\def\slash#1{\setbox0=\hbox{$#1$}               % set a box for #1
   \dimen0=\wd0                                 % and get its size
   \setbox1=\hbox{/} \dimen1=\wd1               % get size of /
   \ifdim\dimen0>\dimen1                        % #1 is bigger
      \rlap{\hbox to \dimen0{\hfil/\hfil}}      % so center / in box
      #1                                        % and print #1
   \else                                        % / is bigger
      \rlap{\hbox to \dimen1{\hfil$#1$\hfil}}   % so center #1
      /                                         % and print /
   \fi}                                         %
\documentstyle[12pt,latexsym]{article}

\begin{document}
\begin{titlepage}
\begin{flushright}
UM-TH-95-06\\
March 1995\\
\end{flushright}
\vskip 1.5cm
\begin{center}
{\large\bf Production Cross-sections for Unstable Particles}
\vskip 1cm
{\large Robin G. Stuart}
\vskip 1cm
{\it Randall Physics Laboratory,\\
 University of Michigan,\\
 Ann Arbor, MI 48109-1120,\\
 USA\\}
\bigskip
and\\
\bigskip
{\it Instituto de F\'\i sica,\\
 Universidad Nacional Aut\'onoma de M\'exico,\\
 Apartado Postal 20-364, 01000 M\'exico D. F.}\\
\end{center}
\vskip .5cm
\begin{abstract}
The top-quark, $W$ and $Z^0$ bosons have widths that are a sizable
fraction of their masses and will be produced copiously at upcoming
accelerators. Yet $S$-matrix theory cannot treat unstable particles as
external states. Dealing with complete matrix elements involving their
decay products complicates calculations considerably and is unnecessary
in many practical situations. It is shown how to construct
physically meaningful production cross-sections for unstable particles
by extracting that part of the matrix element that corresponds to
finite-range
space-time propagation. This procedure avoids the need to define unstable
particles in external states and we argue its favour as
providing a solution to a long-standing problem in physics.
As an example the results are applied to the calculation of the cross-section
$\sigma(e^+e^-\rightarrow Z^0Z^0)$.
\end{abstract}
\vfill
\begin{flushleft}
\end{flushleft}
\end{titlepage}

\setcounter{footnote}{0}
\setcounter{page}{2}
\setcounter{section}{0}
\newpage

\section{Introduction}

The treatment of unstable particles in Quantum Field Theory is fraught
with difficulties. The problems arise from the fact that the particles cannot
be represented as asymptotic states because they decay a
finite distance from the interaction region. Veltman \cite{Veltman}
showed that the $S$-matrix satisfies unitarity and causality
on the Hilbert space of stable particle states even when unstable
particles appeared in intermediate states. Many unstable particles are so
short-lived that they cannot be observed directly and only their decay
products are seen. In such cases the treatment of processes involving
unstable particles in terms of their stable decay products is an accurate
description of the physics.  It does however substantially increase the
complexity of a calculation as the number of external particles that
must be dealt with is larger. Treating unstable particles in this way
also begs the question of their existence. Muons, kaons, $B$-mesons and
neutrons are observed directly and can be manipulated experimentally
before they decay and thus considering them purely in terms of their
decay products is clearly inadequate. Moreover there is no fundamental
difference between these particles and their shorter-lived cousins.
The difference is merely one of experimental resolution.

The problem of the unstable particle was first formulated by Peierls
\cite{Peierls} in the early 1950's and is particularly relevant at the
present time with the upcoming commissioning of LEP200. At this
machine electrons and positrons are collided with sufficient energy
to produce $W^+W^-$ and $Z^0Z^0$ pairs.
Experimental precision is such that
electroweak radiative corrections are required in the comparison of
theory with experiment. However the $W$ and $Z^0$ have a finite width
that is approximately 3\% of their mass and cannot, by any means, be
considered stable. To obtain the required theoretical accuracy the complete
production and decay
process should be considered. This is a process that contains six external
particles and the calculation of the complete electroweak radiative
corrections is daunting indeed. It is certainly very much more complicated
than for the already taxing corrections to $e^+e^-\rightarrow W^+W^-$
\cite{LemoineVeltman,Philippe,Bohmetal,FleiJegeZral}
with only four. In addition, four fermion final states
are produced in $e^+e^-$ collisions via mechanisms other
than intermediate $W$ or $Z^0$ production. All such processes should be
calculated and included for a consistent calculation.

Width effects have also been considered as a possible mechanism
for generating large
$CP$-violating effects in top quark decays \cite{Pilaftsis,EilaHeweSoni}.

In most situations to date the finite lifetime of the unstable
particle is
either ignored or allowed for by what amounts to a convolution of the
stable particle matrix element with a Breit-Wigner distribution
\cite{MutaNajiWaka, DennSack}. The relative merits of a variety
of expansion techniques, that had previously been
used in other physical situations, to treat finite width effects
has also been examined \cite{AeppCuypOlde,AeppOldeWyl}.

A number of problems exist with previous analyses.
The authors of ref.\cite{AeppCuypOlde} apparently misinterpret the method
described in ref.s\cite{Stuart1,Stuart3} as being an expansion
of the matrix element about the real renormalized mass followed
by the addition, by hand, of an imaginary part to that mass.
In fact the method involves Laurent expansion of the
complete matrix element about its complex pole followed by the
perturbative expansion of the pole position, residue and background,
so generated, about the renormalized mass. The error is repeated in
ref.\cite{AeppOldeWyl} where it leads to the appearance of so-called
`threshold singularities' and/or complex scattering angles when one
is below production threshold. This feature is clearly a calculational
artifact. The branch point corresponding to the production of a pair
of unstable $W$ bosons, as considered by these authors, lies away from
the real $s$-axis. One therefore never encounters the branch point
as the centre-of-mass energy changes and one is neither strictly above
nor below threshold. The apparent problem of threshold singularities
arises when one begins an expansion using the real renormalized mass
which places the threshold branch point on the real axis.

The question, ``What is the production cross-section for $Z^0Z^0$
in $e^+e^-$ annihilation?'',
should be a physically well-posed one. One need only perform a
{\it gedanken} weakening of the Standard Model couplings
of the $Z^0$'s in order
to make them sufficiently long-lived that their existence could be
confirmed by, say, vertex detectors. Above their production threshold,
physical $Z^0$ pairs would be expected to be the dominant source of
$(f_1\bar f_1)(f_2\bar f_2)$ final states because of their resonant
enhancement and thus a
definitive answer would avoid the need to consider processes with
six external particles. Yet $S$-matrix theory provides no unambiguous
response.

Note in passing that for the inverse process, the question as to
what is the lifetime of an unstable particle is not well-posed. It
has been known for a long time \cite{Schwinger} that this depends
on the manner in which the particle was prepared. It is, for example,
meaningful to inquire as to the lifetime of an unstable particle
with a definite 4-momentum.

Many years ago, a number of authors
\cite{Stapp,Gunson} used residues at poles in the $S$-matrix element
for processes containing intermediate unstable particles
to define $S$-matrix elements for external unstable
particles. Although these generalized $S$-matrix elements
possess intriguing mathematical properties, such as generalized
unitarity relations, their physical meaning is
less than clear and it is not obvious how they may be used to
calculate measurable production cross-sections.

In this paper an answer will be given to the question posed above.
Although standard perturbation theory cannot tolerate unstable
particles as asymptotic states, it will be shown that it is
nevertheless possible to formulate  a physically motivated
procedure to consistently extract from the $S$-matrix
element for $e^+e^-\rightarrow (f_1\bar f_1) (f_2\bar f_2)$
that part which proceeds via the production of an intermediate
$Z^0$ pair and their subsequent decay. Summation over all such
final states then gives the $Z^0$ production cross-section. In
this procedure standard perturbation theory is employed. We do
not attempt to define an effective propagator for an unstable
particle or modify the usual Feynman rules in any way. Nor is
an $S$-matrix element defined with external unstable particles
but nevertheless a meaningful physical cross-section is
obtained. The uniqueness of the procedure is discussed at the end
of section 2.

An immediate consequence of this is that the above reaction may
be separated into four distinct physical, and consequently separately
gauge-invariant, processes
\[
\begin{array}{cl}
e^+e^-\longrightarrow&\left\{ \begin{array}{l}
           Z^0\rightarrow f_1\bar f_1\\
           Z^0\rightarrow f_2\bar f_2
           \end{array}\right.\\
 & \\
e^+e^-\longrightarrow&\left\{ \begin{array}{l}
           Z^0\rightarrow f_1\bar f_1\\
           f_2\bar f_2
           \end{array}\right.\\
 & \\
e^+e^-\longrightarrow&\left\{ \begin{array}{l}
           f_1\bar f_1\\
           Z^0\rightarrow f_2\bar f_2
           \end{array}\right.\\
 & \\
e^+e^-\longrightarrow&\left\{ \begin{array}{l}
           f_1\bar f_1\\
           f_2\bar f_2
           \end{array}\right.
\end{array}\]
The first of these is the one we will consider here.
It is the dominant mode in a variety of kinematic regions of
experimental interest. The others may be calculated by similar
methods if the experimental situation requires it.

The calculation of $\sigma(e^+e^-\rightarrow Z^0 Z^0)$ is technically
much more demanding than for the process
$e^+e^-\rightarrow Z^0\rightarrow f\bar f$.
In the latter case the $Z^0$ is produced with a fixed invariant mass and
but for the former the invariant masses of
the $Z^0$'s vary over some kinematically allowed region and
must somehow be incorporated into phase-space integrations.
As discussed above, na\"\i ve
generalizations of the methods of ref.s\cite{Stuart1,Stuart3} soon
encounter difficulties or apparent arbitrariness in how to proceed.
It will be shown that, by carefully respecting the Lorentz-structure
of the analytic $S$-matrix at each stage, a unique procedure is
indicated with no flexibility in the final result.

A number of authors \cite{MelnYako1, MelnYako2, Khoze} have studied the
effects of interactions between the final-state decay products of unstable
particles. These are generally found to be suppressed and the physical
reason for this will be given at the end of the paper.

\section{Identification of an Unstable Particle}

We will begin by examining the form and properties of the
dressed propagator or Green's function for an unstable particle.
For simplicity a scalar particle will be considered but the generalization
to fermions or vector bosons is straightforward.
Although the dressed propagator does not by itself represent a physical
observable we will use it
to investigate what properties of the analytic $S$-matrix
element may be employed to indicate the presence of an unstable particle.

The equation of motion of a free scalar particle are given by the
Klein-Gordon equation
\begin{equation}
(\Box+m^2)\phi(x)=0.
\end{equation}
The space-time evolution of the scalar field is governed by the
propagator $\Delta(x^\prime-x)$ which in coordinate space
may be represented as
\begin{equation}
\Delta(x^\prime-x)=\int \frac{d^4k}{(2\pi)^4}
                   \frac{e^{-ik\cdot(x^\prime-x)}}{k^2-m^2+i\epsilon}.
\end{equation}
When interactions are switched on the dressed propagator
\begin{equation}
\Delta(x^\prime-x)=\int \frac{d^4k}{(2\pi)^4}
                   \frac{e^{-ik\cdot(x^\prime-x)}}
                                 {k^2-m^2-\Pi(k^2)+i\epsilon}
\end{equation}
is generated. Here $\Pi(k^2)$ is scalar's self-energy.
For reasons that will become apparent soon the integrand in
will be split into three pieces. Thus
\begin{eqnarray}
\Delta(x^\prime-x)&=&\int \frac{d^4k}{(2\pi)^4}
        e^{-ik\cdot(x^\prime-x)}\nonumber\\
 &\times&\left[\frac{1}{2k_0}.\frac{F(s_p)}{k_0-\sqrt{\vec{k}^2+s_p}}
         +\frac{F(k^2)-F(s_p)}{k^2-s_p}
         +\frac{1}{2k_0}.\frac{F(s_p)}{k_0+\sqrt{\vec{k}^2+s_p}}\right].
\label{eq:splitprop}
\end{eqnarray}
The quantity $s_p$ is the position of the complex pole of
the dressed propagator. It is a solution to the equation
\begin{equation} s-m^2+\Pi(s)=0 \end{equation}
Also
\begin{equation}
F(s)=\frac{s-s_p}{s-m^2+\Pi(s)}\label{eq:FZZ}
\end{equation}
from which it follows by l'H\^opital's rule
$F(s_p)=\left(1+\Pi^\prime(s_p)\right)^{-1}$.

Performing the integration in $k_0$ on the first and third terms in
(\ref{eq:splitprop}) for $t\ne t^\prime$ gives
\begin{eqnarray}
\Delta(x^\prime-x)&=&-i\int \frac{d^3k}{(2\pi)^3 2k_0}
                    e^{-ik\cdot(x^\prime-x)}\theta(t^\prime-t)F(s_p)
                     \nonumber\\
                  & &+\int\frac{d^4k}{(2\pi)^4}\frac{F(k^2)-F(s_p)}{k^2-s_p}
                     \nonumber\\
                  & &-i\int \frac{d^3k}{(2\pi)^3 2k_0}
                    e^{ik\cdot(x^\prime-x)}\theta(t-t^\prime)F(s_p).
\label{eq:intsplitprop}
\end{eqnarray}
In eq.(\ref{eq:intsplitprop}) $k_0=\sqrt{\vec{k}^2+s_p}$.
The second term on the right-hand side of
(\ref{eq:splitprop}) has no poles in $k_0$ near the physical region and
is identifiable as a contact interaction locally.

The first terms of (\ref{eq:splitprop}) and (\ref{eq:intsplitprop})
describe the propagation of the scalar field forward in time. The
last term is usually thought of as the forward propagation of the
anti\-part\-icle. It is tempting to conclude that the first term
describes the motion of the physical unstable particle.
Indeed terms of rather similar form are obtained when
Bogoliubov causality is applied to the dressed propagator \cite{Diagrammar}.
The problem with the decomposition in (\ref{eq:splitprop})
is that the first and last terms are not separately Lorentz invariant.
One would expect that all observers should agree on whether an
unstable particle has been created. In processes in which
the propagator is connected to external stable particles, one or other
of the particle or anti\-part\-icle pieces will be energetically disfavoured
and represent a virtual state that is protected from observation by the
uncertainty principle. Both pieces must be taken together to obtain a
Lorentz-invariant expression. The propagator (\ref{eq:splitprop}) may
then be written
\begin{equation}
\Delta(x^\prime-x)=\int\frac{d^4k}{(2\pi)^4} e^{-ik\cdot(x^\prime-x)}
 \left[\frac{F(s_p)}{k^2-s_p}+\frac{F(k^2)-F(s_p)}{k^2-s_p}\right].
\end{equation}
The first term is the finite-range piece and the second is a contact term.

In the following we will identify the unstable particle by its
finite-range propagation within a physical matrix element which may be
obtained by procedures similar to those applied above to the dressed
propagator. This characterization of an unstable particle avoids the need
define them in external states and is therefore amenable to $S$-matrix
theory. It is as equally applicable to the directly-observable
neutron as to the $Z^0$.

We will be interested in calculating production
cross-sections for unstable particles and will therefore drop contact
terms since they describe the prompt production of the final state
without the intermediate production of an unstable particle.
The contact term is normally subdominant and in many practical situations
lies below experimental error. In such cases it may also be dropped and
the production cross-section for the unstable particle provides an
adequate description of the physics. Where the contact term is required,
as for example in LEP1 physics,
it may be added without difficulty \cite{Stuart1}.

It is interesting to note that the finite-range and contact interaction
are both produced by the same field yet are physically distinguishable.
There is therefore not a one-to-one correspondence between the field
and the particle.

As stated above we have chosen to tag the existence of an unstable particle
by its finite space-time propagation which involves decomposition of an
$S$-matrix element into two distinguishable pieces. It may be one could
use some other property to identify the presence of an unstable particle
and divide up the matrix element accordingly but such an alternative
would necessarily mean mixing propagating and non-propagating modes of
the field in which case the physical interpretation becomes unclear.
Identifying the unstable particle by its finite space-time propagation
seems, however, to be the choice which is most physically and intuitively
reasonable. It is, after all, this property that allows $b$-quarks to be
identified by vertex detectors.

\section{The cross-section for $e^+e^-\rightarrow Z^0 Z^0$}

The above considerations allow the cross-section for
$e^+e^-\rightarrow Z^0 Z^0$ to be calculated. We start from the
the full matrix element for
$e^+e^-\rightarrow(f_1\bar f_1)(f_2\bar f_2)$.
For the present purposes it will be assumed that the fermions are massless.

The calculation of the production cross-section for $Z^0$ pairs
proceeds along classical lines using standard Feynman rules.
The squared invariant masses of the $f_1\bar f_1$ and $f_2\bar f_2$ pairs
are $p_1^2$ and $p_2^2$.
The matrix element that accounts for physical
$Z^0Z^0$ production is obtained by extracting the part that is resonant
in both $p_1^2$ and $p_2^2$. The cross-section is then
obtained by squaring and summing over all possible final states.

The part of the matrix element that is resonant in $p_1^2$
but not $p_2^2$
describes the production of a physical $Z^0$ and a pair of fermions
$f_2\bar f_2$. This piece must be combined with other Feynman diagrams
yielding the same final state but no additional intermediate $Z^0$
in order to produce a gauge-invariant result.

The part of the matrix element that is resonant in $p_2^2$
describes $Z^0$ production in association with the fermions
$f_1\bar f_1$ and the part that is resonant in neither variable
describes direct four-fermion production.
Thus the complete matrix element $Z^0$ separates into distinct pieces
corresponding to the four distinct physical processes given previously.

The part of the full matrix element that can give rise to doubly
resonant contributions can be written as
\begin{eqnarray}
{\cal M}&=&\sum_i [\bar v_{e^+} T^i_{\mu\nu} u_{e^-}]
M_i(t,u,p_1^2,p_2^2)\nonumber\\
& &\ \ \ \ \ \ \ \times\frac{1}{p_1^2-M_Z^2-\Pi_{ZZ}(p_1^2)}
[\bar u_{f_1}\gamma^\mu(V_{Zf_L}(p_1^2)\gamma_L
                       +V_{Zf_R}(p_1^2)\gamma_R) v_{\bar f_1}]
\label{eq:fullZZ}\\
& &\ \ \ \ \ \ \ \times\frac{1}{p^2_2-M_Z^2-\Pi_{ZZ}(p_2^2)}
[\bar u_{f_2}\gamma^\nu(V_{Zf_L}(p_2^2)\gamma_L
                       +V_{Zf_R}(p_2^2)\gamma_R) v_{\bar f_2}]
\nonumber
\end{eqnarray}
where $T_{\mu\nu}^i$ are Lorentz covariant tensors that span the tensor
structure of the matrix element and $\gamma_L$, $\gamma_R$ are the
usual helicity projection operators. The $M_i$, $\Pi_{ZZ}$ and $V_{Zf}$
are Lorentz scalars that are analytic functions of the
independent Lorentz invariants of the problem. It is always possible to
separate the matrix element into so-called `standard' Lorentz covariants
and Lorentz scalar functions \cite{Hearn,Hepp1,Williams,Hepp2}.
There is a limited amount of flexibility as to where this
separation is made. One must not, for example, introduce spurious kinematic
singularities and hence the final results for the matrix elements
and cross-sections obtained in what follows are not affected by
this freedom. It is sometimes helpful, although not
essential, that the $T_{\mu\nu}^i$ be constructed to be separately
gauge-invariant \cite{BardeenTung}. As before the external fermion
wave-functions are denoted $u$ and $v$ with appropriate subscripts.
The $Z$-$\gamma$ mixing has been neglected but it will
not alter the structure of the final result \cite{Stuart3}.

To extract the piece of the matrix element that corresponds to finite
propagation of both $Z^0$'s we extract the leading term in a Laurent
expansion in $p_1^2$ and $p_2^2$ of the analytic Lorentz-invariant
part of eq.(\ref{eq:fullZZ}) leaving the Lorentz-covariant part
untouched. This is the doubly-resonant term and is given by
\begin{eqnarray}
{\cal M}&=&\sum_i [\bar v_{e^+} T^i_{\mu\nu} u_{e^-}]
M_i(t,u,s_p,s_p)\nonumber\\
& &\ \ \ \ \ \ \ \times\frac{F_{ZZ}(s_p)}{p_1^2-s_p}
[\bar u_{f_1}\gamma^\mu(V_{Zf_L}(s_p)\gamma_L
                       +V_{Zf_R}(s_p)\gamma_R) v_{\bar f_1}]
\label{eq:resZZ}\\
& &\ \ \ \ \ \ \ \times\frac{F_{ZZ}(s_p)}{p_2^2-s_p}
[\bar u_{f_2}\gamma^\nu(V_{Zf_L}(s_p)\gamma_L
                       +V_{Zf_R}(s_p)\gamma_R) v_{\bar f_2}]
\nonumber
\end{eqnarray}
where $F_{ZZ}$ defined by a relation like (\ref{eq:FZZ}).
It should be emphasized that eq.(\ref{eq:resZZ}) the exact form
of the doubly-resonant matrix element to all orders in perturbation
theory that we will now specialize to leading order.
It is free of threshold singularities noted that were found by other
authors \cite{AeppOldeWyl}.
Although $M_i$ was obtained directly by means of eq.(\ref{eq:fullZZ})
the general from of eq.(\ref{eq:resZZ}) with its factorization between
initial and final states is guaranteed by Fredholm theory.
Note that the external wave-functions do not
figure in the expansion as they are external sources and sinks and
do not contribute to defining the finite-range part.
The kinematic factors associated with external particles are thus
unchanged by the expansion so that squaring of the matrix element and
phase space integrals can be performed without added difficulty.
Moreover since the doubly-resonant matrix element is independent of the
singly- and non-resonant parts of the complete matrix element,
(\ref{eq:resZZ}), enjoys the same invariance properties, such as
gauge-independence, as the complete matrix element.

A word is in order here about momentum conservation within the Lorentz scalar
functions such, as $M_i$, during analytic continuation.
The analytic continuation of the $M_i$ to complex values of $p_1^2$
and $p_2^2$ is a well-defined procedure that formed the basis for defining
generalized $S$-matrix elements for external unstable particles and about
which an extensive literature exists \cite{Stapp,Gunson}.
The $S$-matrix requires that the external massless
fermions stay on-shell and conservation of 4-momentum implies
that ${\cal M}$ can be a function of at most four independent Lorentz
scalars \cite[section 4.3]{AnalyticSMat}.
These we are free to take as $t$, $u$, $p_1^2$ and $p_2^2$. Any of
these variables may be continued without upsetting the constraints of momentum
conservation. Thus for example, the well-known condition $s+t+u=p_1^2+p_2^2$
remains satisfied as $p_1^2$ and $p_2^2$ are continued away from the real
axis by compensating changes in the dependent variable $s$.

In lowest order eq.(\ref{eq:resZZ}) becomes, up to overall multiplicative
factors,
\begin{eqnarray}
{\cal M}&=&\sum_{i=1}^2 [\bar v_{e^+} T^i_{\mu\nu} u_{e^-}]M_i\nonumber\\
& &\ \ \ \ \times\frac{1}{p_1^2-s_p}
[\bar u_{f_1}\gamma^\mu(V_{Zf_L}\gamma_L+V_{Zf_R}\gamma_R)v_{\bar f_1}]\\
& &\ \ \ \ \times\frac{1}{p_2^2-s_p}
[\bar u_{f_2}\gamma^\nu(V_{Zf_L}\gamma_L+V_{Zf_R}\gamma_R)v_{\bar f_2}].
\nonumber
\label{eq:lowestZZ}
\end{eqnarray}
where
$T^1_{\mu\nu}=\gamma_\mu(\slash{p}_{e^-}-\slash{p}_1)\gamma_\nu$,
$M_1=t^{-1}$;
$T^2_{\mu\nu}=\gamma_\nu(\slash{p}_{e^-}-\slash{p}_2)\gamma_\mu$,
$M_2=u^{-1}$
and the final state vertex corrections take the form
$V_{Zf_L}=ie\beta_L^f\gamma_L$ and $V_{Zf_R}=ie\beta_R^f\gamma_R$,
The left- and right-handed couplings of the $Z^0$ to a fermion $f$ are
\[
\beta_L^f=\frac{t_3^f-\sin^2\theta_W Q^f}{\sin\theta_W\cos\theta_W},
\hbox to 2cm{}
\beta_R^f=-\frac{\sin\theta_W Q^f}{\cos\theta_W}.
\]

Squaring the matrix element and integrating over the final state
momenta for fixed $p_1^2$ and $p_2^2$ gives
\begin{equation}
\frac{\partial^3\sigma}{\partial t\,\partial p_1^2\,\partial p_2^2}
                   =\frac{\pi\alpha^2}{s^2}
                    (\vert\beta_L^e\vert^4+\vert\beta_R^e\vert^4)
                     \left\{\frac{t}{u}+\frac{u}{t}
                     +\frac{2(p_1^2+p_2^2)^2}{ut}
                     -p_1^2p_2^2\left(\frac{1}{t^2}+\frac{1}{u^2}\right)\right\}
\ \rho(p_1^2)\ \rho(p_2^2)
\label{eq:diffxsec}\end{equation}
with
\begin{eqnarray*}
\rho(p^2)&=&\frac{\alpha}{6\pi}
    \sum_f(\vert\beta_L^f\vert^2+\vert\beta_R^f\vert^2)
          \frac{p^2}{\vert p^2-s_p\vert^2}
               \theta(p_0)\theta(p^2)\\
 &\approx&\frac{1}{\pi}
         .\frac{p^2 (\Gamma_Z/M_Z)}{(p^2-M_Z^2)^2+\Gamma_Z^2 M_Z^2}
          \theta(p_0)\theta(p^2)
\end{eqnarray*}
Note that $\rho(p^2)\rightarrow \delta(p^2-M_Z^2)\theta(p_0)$
as $\Im(s_p)\rightarrow 0$ which is the result obtained by cutting a
free propagator. The variables $s$, $t$, $u$, $p_1^2$ and $p_2$ in
eq.(\ref{eq:diffxsec}) arise from products of standard covariants and external
wave functions and therefore take real values dictated by the kinematics.

Integrating over $t$, $p_1^2$ and $p_2^2$ leads to
\begin{equation}
\sigma(s)=\int_0^s dp_1^2
          \int_0^{(\sqrt{s}-\sqrt{p_1^2})^2} dp_2^2
          \sigma(s;p_1^2,p_2^2)\ \rho(p_1^2)\ \rho(p_2^2),
\label{eq:ZZxsec}
\end{equation}
where
\[
\sigma(s;p_1^2,p_2^2)=\frac{2\pi\alpha^2}{s^2}
       (\vert\beta_L^e\vert^4+\vert\beta_R^e\vert^4)
       \left\{\left(\frac{1+(p_1^2+p_2^2)^2/s^2}
{1-(p_1^2+p_2^2)/s}\right)\ln\left(\frac{-s+p_1^2+p_2^2+\lambda}
                                    {-s+p_1^2+p_2^2-\lambda}\right)
-\frac{\lambda}{s}\right\}
\]
and $\lambda=\sqrt{s^2+p_1^4+p_2^4-2sp_1^2-2sp_2^2-2p_1^2p_2^2}$.
For $p_1^2=p_2^2=M_Z^2$ this agrees with known results \cite{Brown}.

The form of eq.(\ref{eq:ZZxsec}) superficially rather similar to
expressions obtained by others \cite{MutaNajiWaka,DennSack}.
These authors apply their method to tree-level matrix elements to obtain
a cross-section in the form of a gauge-invariant convolution integral.
Applying the same method to the one-loop or higher matrix-elements
would require the evaluation the form factors, $M_i$,
cross-section in the integrand with off-shell momenta, $p_1^2$ and $p_2^2$,
and will thus generate a gauge-dependent result.
Here the extraction of the full doubly resonant part from the
matrix-element generates a gauge-invariant result at all orders.
In higher orders the integrations in $p_1^2$ and $p_2^2$ are relatively
simple as they only need to be performed over kinematically
generated factors and not over complete higher order form-factors.
The price is that the form-factors must be evaluated for complex arguments.

\section{Discussion}

In the recent past attempts have been made to produce
gauge-invariant self-energies, vertex corrections and box diagrams
in order to ease problems of spurious gauge-dependence that can often arise
in certain types of calculations. This is done by extracting `universal'
pieces from vertex and box diagrams and combining with self-energies.
The universal pieces have been obtained by a variety of methods
\cite{KennedyLynn,KLIS,DegrassiSirlin}
such as the so-called pinch technique. These
self-energies may be resummed in the usual way along
with their universal pieces. While there is no doubt that the self-energies
do not display explicit gauge parameter dependence,
their use in resummations is highly questionable
as there are no Feynman diagrams that
would generate such a resummation of vertex pieces. Adding universal
pieces to the self-energies in this way also runs the risk, as
was pointed out in ref.\cite{Stuart3}, of shifting
the position of the complex pole of the $S$-matrix element, that is
governed solely by the self-energy. The residue at the pole is also
at risk of being altered by these procedures.
Although there are indications that this does not happen at one loop,
the pinch technique provides no firm proof that this feature
persists in higher orders.

In ref.s\cite{Stuart1,Stuart3} it was shown how to obtain an exactly
gauge-invariant perturbation expansion to all orders
of the matrix element for
the process $e^+e^-\rightarrow f\bar f$ near the $Z^0$ resonance.
This involved performing a Laurent expansion of the matrix element
about the pole and thereby extracting the resonant and non-resonant
background terms. Gauge invariance of the non-resonant background
is achieved by a combination of non-resonant Feynman diagrams
and non-resonant contributions coming from the dressed self energy.
In the alternative approaches discussed above, quite the opposite is
true. There pieces from vertices and boxes are migrated into the
self-energy. Yet the split between resonant and non-resonant terms
is not just a mathematical trick. As seen in the foregoing, resonant
and non-resonant parts describe distinct physical processes.

Recently the background field method has been used to generate
gauge-invariant self-energies \cite{DennWeigDitt}.
This constitutes a genuine
self-consistent perturbation expansion that can, in principle, be
extended to all orders. Resummations can be performed without the
risk of shifting the pole position. Nevertheless, the resulting
matrix elements should be subjected to Laurent expansion in order to
make a clean separation between distinct physical processes that is a
feature of the exact $S$-matrix element.

A number of groups \cite{MelnYako1, MelnYako2, Khoze} have
examined the production of charged unstable particles; $W^+W^-$,
$t\bar t$ etc. Many such effects are suppressed because the
unstable particle propagates a finite distance from the production
point before producing the final-state decay products.

The phenomenon of suppression of corrections in the presence of
resonant states has already been observed for the
case of the $Z^0$-boson \cite{KuhnStuart,KuhnJadaStuaWas}.
In an exact calculation of the QED
initial-final state interference on the resonant part of the
matrix element for $e^+e^-\rightarrow f\bar f$. It was found that
the corrections pass through zero near $s=M_Z^2$ which may be
understood as the $Z^0$ propagating a finite distance from its
production point before decaying.

\section{Acknowledgments}
The author wishes to thank D.\ Williams for many useful discussions.
This work was supported in part by the U.S. Department of Energy.

\end{document}